\documentclass[]{aastex631}

\submitjournal{JCAP}
\shorttitle{New Evidence Quasar ...}
\shortauthors{Slagter R J}
\graphicspath{{./}{figures/}}
\begin{document}
\title{New Evidence of the Azimuthal Alignment of Quasars Spin Vector in the LQG U1.28, U1.27, U1.11, Cosmologically  Explained.}
\correspondingauthor{Reinoud Jan Slagter}
\email{info@asfyon.com}
\author[0000-0002-2429-9866]{Reinoud J. Slagter}
\affiliation{Astronomisch Fysisch Onderzoek Nederland (ASFYON)\\ and\\ 
University of Amsterdam (on leave) \\
1405EP Bussum, The Netherlands}
\begin{abstract}
There has been observational evidence about  spin axes of quasars in large quasar groups  correlated over  hundreds of Mpc. This is seen  in the radio spectrum as well as in the optical range. There is not yet a satisfactory explanation of this "spooky" alignment.
This alignment cannot be explained by mutual interaction at the time that quasars manifest themselves optically. A cosmological explanation  could be possible in the formation of superconducting vortices (cosmic strings) in the early universe, just after the symmetry-breaking phase of the universe.
We gathered  from the NASA/IPAC  and SIMBAD  extragalactic databases  the right ascension, declination, inclination, position angle and eccentricity of the host galaxies of 3 large quasar groups  to obtain the azimuthal and polar angle of the spin vectors.
The alignment of the azimuthal angle of the spin vectors of quasars in their host galaxy is confirmed in the large quasar group U1.27 and compared with two other groups in the vicinity, i.e., U1.11 and U1.28, investigated by Clowes (2013).
It is well possible that the azimuthal angle alignment fits the predicted azimuthal angle dependency in the theoretical model of the formation of general relativistic superconducting vortices, where the initial axially symmetry is broken  just after the symmetry breaking of the scalar-gauge field.
\end{abstract}
\section{Introduction} \label{sec:intro}
A large quasar group (LQG) is a cluster of quasars that makes the largest astronomical structures in the current universe. Their  sizes can be of the order of hundreds of Mpc. 
Astronomers believe that a quasar is an active galactic nuclei (AGN) with a vibrant  eruption of radiation both optical and in radio range originated by a spinning (Kerr-) black hole, surrounded by an accretion disk.
According to Taylor and Jagannathan \citep{Taylor2016}, a LQG has an internal non-uniform distribution of spin vectors seen in the radio spectrum and the optical spectrum as observed by Hutsemekers et al.\citep{Hutsemekers2014}.
This coherence is mysterious and cannot be explained by mutual interaction at the time scale of primordial galaxies formation but rather by use of a more advanced method \citep{Slagter2018}.
In a recent  study, Slagter \citep{Slagter2021} found that the azimuthal angle of the spin vector of quasars in their host galaxies in six  quasar groups, show preferred directions. 
This is demonstrated through  an emergent azimuthal angle dependency of the general relativistic Nielsen-Olesen (NO) vortices at the point after the symmetry breaking at grand unified theory (GUT)-scale.
This review focuses on  three more other LQG, studied by Clowes \citep{Clowes2012,Clowes2013}.
\section{Results}\label{sec:results}
From the NASA/IPAC extragalactic database and SIMBAD  we extract for the three LQG U1.11, U1.27 and U.28 the right ascension, declination, inclination, position angle and eccentricity of the host galaxies. The 3-D orientation of the spin vectors can then be calculated\citep{Pajowska2019}. In figure 1 and 2 we plotted the azimuthal angle. Without statistical analysis one can conclude that the preferred orientations are evident. In the case of LQG U1.27 (see table 1 and 2 for the data), we fitted two trigonometric functions on the distribution, which can theoretically be explained (section 3).
\begin{figure}[h]
\centerline{\includegraphics[scale=.85]{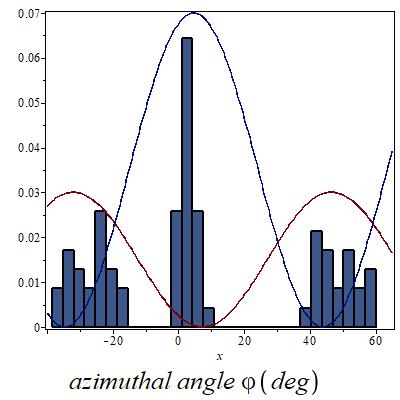}}
\caption{{\it Plot of the azimuthal angle $\varphi$ in degrees. This shows the distribution of the azimuthal angle  of the spin vectors in the LQG  U1.27 (N=71) with a best-fit of two trigonometric  functions with a phase shift range of $45^o$.}}
\end{figure}
\begin{figure}
\centerline{\includegraphics[scale=.6]{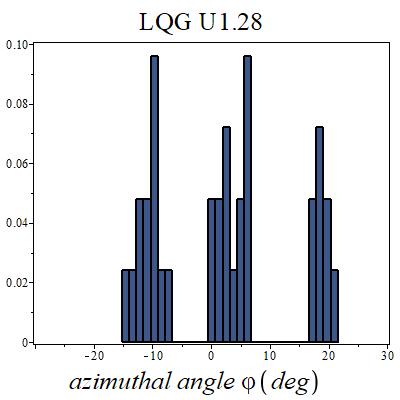}
\includegraphics[scale=.6]{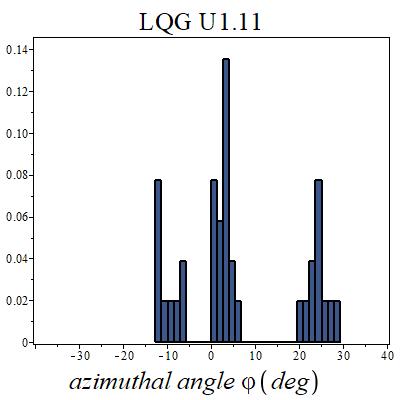}}
\caption{{\it Plot showing the distribution of the azimuthal angle $\varphi$ for the LQG U1.28 (N=34) and U1.11 (N=38).}}
\end{figure}
\section{The theoretical model}\label{sec:theory}
A linear approximation of wavelike solutions of the Einstein equations is not adequate when one is dealing with high curvature (or high energy scale), i.e., close to the horizons of black holes or in the early stage of the universe at the time of  mass formation by the Higgs mechanism. There will be a "back-reaction" on the background spacetime.
There is a powerful approximation method which can deal with these non linearities: the multiple-scale method. Pioneer work was done by ChoquetBruhat \citep{ChoquetBruhat1968}.
One expands the relevant fields \citep{Slagter1986}
\begin{equation}
V_i=\sum_{n=0}^{\infty}\frac{1}{\omega^n}F_i^{(n)}({\bf x},\xi),\label{3-1}
\end{equation}
where $\omega$ represents a dimensionless  parameter ("frequency"), which will be large. Further, $\xi=\omega\Theta({\bf x})$, with $\Theta$ a scalar (phase) function on the manifold.
The small parameter $\frac{1}{\omega}$ can also be the ratio of the characteristic wavelength of the perturbation to the characteristic dimension of the background. On warped spacetimes it could also be the ratio of the extra dimension to the background dimension.
In the vacuum case, we expand the metric
\begin{equation}
g_{\mu\nu}=\bar g_{\mu\nu}+\frac{1}{\omega}h_{\mu\nu}({\bf x},\xi)+\frac{1}{\omega^2}k_{\mu\nu}({\bf x},\xi)+ ...,\label{3-2}
\end{equation}
where we defined
\begin{eqnarray}
\frac{dg_{\mu\nu}}{dx^\sigma}=g_{\mu\nu,\sigma}+\omega l_\sigma\dot g_{\mu\nu},\quad
g_{\mu\nu,\sigma}=\frac{\partial g_{\mu\nu}}{\partial x^\sigma},\quad 
\dot g_{\mu\nu}=\frac{\partial g_{\mu\nu}}{\partial \xi},\label{3-3}
\end{eqnarray}
with $l_\mu =\frac{\partial\Theta}{x^\mu}$.
One then says that
\begin{equation}
V_i=\sum_{n=-m}^{\infty}\frac{1}{\omega^n}F_i^{(n)}({\bf x},\xi)\label{3-4}
\end{equation}
is  an approximate wavelike solution of order n of the field equation, if $F_i^{(n)}=0, \forall n$. One can substitute the expansion into the field equations.
The Ricci tensor then expands as
\begin{equation}
R_{\mu\nu}\rightarrow \omega R_{\mu\nu}^{(-1)}+\Bigl(\bar R_{\mu\nu}+R_{\mu\nu}^{(0)}\Bigr)+\frac{1}{\omega}R_{\mu\nu}^{(1)} + ...\label{3-5}
\end{equation}
By equating the subsequent orders to zero, we obtain
\begin{equation}
R_{\mu\nu}^{(-1)}=0=\frac{1}{2}\bar g^{\beta\lambda}(l_\lambda l_\mu\ddot h_{\beta\nu}+l_\nu l_\beta\ddot h_{\mu\lambda}-l_\lambda l_\beta\ddot h_{\mu\nu}-l_\nu l_\mu\ddot h_{\beta\lambda}),\label{3-6}
\end{equation}
\begin{eqnarray}
R_{\mu\nu}^{(0)}+\bar R_{\mu\nu}=0,\qquad  R_{\mu\nu}^{(1)}=0,\qquad ....\label{3-7}
\end{eqnarray}
Here we used $l_\mu l^\mu =0$. The rapid variation is observed  in the direction of $l_\mu$. In the radiative outgoing Eddington-Finkelstein coordinates, we have $x^1=u=\Theta({\bf x})=t-r$ and $ l_\mu =(1,0,0,0)$, while the bar stands for the background.
\subsection{Formation of vortices}
In a recent study \citep{Slagter2016,Slagter2017,Slagter2021} we applied this non-linear approximation scheme on a FLRW spacetime. We considered the matter contribution of a gauged complex scalar (Higgs) field. 
Physicists are now convinced that this field plays a fundamental role in the early universe and is responsible for the symmetry breaking in the Standard Model of particle physics. The experimental verification came by the recently observed Higgs particle at CERN. The same field has lived up to its reputation in superconductivity, where the field act as an order parameter to describe the formation of Cooper pairs. 
The scalar field is combined with a gauge field, parameterized as $\Phi=\eta X({\bf x})e^{in\varphi}$ and $A_\mu({\bf x})=\frac{n}{e}(P({\bf x})-1)\nabla_\mu\varphi$,
with $n$ the topological charge or winding number. The trapped  flux of the vortex is expressed as $n\frac{2\pi h}{e}$. The formation of a lattice of quantized magnetic flux tubes was first observed by  Abrikosov \citep{Abrikosov1957} and are described by the famous equations of Ginzberg \citep{Ginzberg1950}. In these models , one needs the quartic potential of the Higgs field, i.e., $V(\Phi)=\frac{1}{8}\lambda\Bigl(\Phi\Phi^*-\eta^2\Bigr)^2$, with $\eta$ the vacuum expectation value. Further, $\frac{m_\Phi}{m_A}=\frac{e^2}{ \lambda}$ is the ratio of the scalar to gauge masses. This potential leads to a nonzero $\eta$ and spontaneous breaking of the U(1) symmetry (note that the parameters are in general temperature dependent). 
The forces existing between the vortices are  the electromagnetic and scalar force. When the vortices get close together, the problem becomes non-linear and the resulting  force depends on the ratio $\frac{e^2}{\lambda}$. For details, see, for example, the text books of Felsager \citep{Felsager1998} and Weinberg \citep{Weinberg2012}.
Moreover, when the vortices are formed in the early stage of the universe, then gravity will come into play. These general relativistic vortex solutions are known as "cosmic strings" \citep{Garfinkle1985, Vilenkin1994}, when they extend to cosmological dimensions. They could possible explain the observed void and filament structures in the universe. 
We expand the scalar and gauge field to second order as
\begin{eqnarray}
A_\mu=\bar A_\mu ({\bf x})+\frac{1}{\omega}B_\mu ({\bf x},\xi) +\frac{1}{\omega^2}C_\mu ({\bf x},\xi) +... ,\label{3-8}
\end{eqnarray}
\begin{eqnarray}
\Phi=\bar\Phi({\bf x}) +\frac{1}{\omega}\Psi({\bf x}, \xi)+\frac{1}{\omega^2}\Xi({\bf x}, \xi)+...,\label{3-9}
\end{eqnarray}
where we write the subsequent orders of the scalar field as
\begin{eqnarray}
\bar\Phi =\eta \bar X(t,r) e^{i n_1 \varphi},\quad \Psi = Y(t,r,\xi) e^{i n_2 \varphi},\quad\Xi = Z(t,r,\xi) e^{i n_3 \varphi}\label{3-10},
\end{eqnarray}
with $n_i$ the winding numbers. 
\subsection{The azimuthal angle dependency: breaking the axial symmetry}
The azimuthal angle $\varphi$ does not reach the partial differential equations (PDE) in the unperturbed case. By quantum fluctuations, the vortex excite in higher $n$-state and will dissociate into $n$ well separated $n=1$ vortices\footnote{The stability of the configuration depends on parameter $\lambda$\citep{Weinberg2012}}, because the energy of the configuration is proportional with $n^2$. 
The topological characterization is a set of isolated points  with winding numbers $n_i$ (the zeros of $\Phi$), with $ n=n_1, n_2, ...$. This n-vortices solution represents a finite energy configuration. 
However, an imprint will be left over of the azimuthal dependency of the orientation of the clustering of Abrikosov vortices lattice in the general relativistic situation.
So the axial symmetry is dynamically broken. 
The azimuthal dependency emerge already to first order in the approximation. For example, the energy-momentum tensor $\bar T_{t\varphi}=0$, while the first order perturbation becomes
\begin{equation}
T_{t\varphi}^{(0)}=\bar X\bar P\dot Yn_1\sin(n_2-n_1)\varphi\label{3-11}
\end{equation}
However, in $T_{t\varphi}^{(1)}$ there appears terms like $\cos(n_2-n_1)\varphi$ and $\sin(n_3-n_1)\varphi$.
The perturbative appearance of a nonzero energy-momentum component $T_{t\varphi}$ can be compared with the phenomenon of bifurcation along the Maclaurin-Jacobi sequence of equilibrium ellipsoids of self-gravitating compact objects, signalling the onset of secular instabilities\citep{GondekRosinska2002}. This shows a similarity with the Goldstone-boson modes of spontaneously broken symmetries of continuous groups. The recovery of the SO(2) symmetry from the equatorial eccentricity takes place at a time comparable to the emission of gravitational waves.

The particular ellipsoid orientation in the frame $(r,\varphi,z)$ expressed as  $\varphi_0\equiv\varphi(t_0)$, is at $t>t_0$ and  determined by the transformation $\varphi \rightarrow \varphi_0- Jt$, where $J$ is the rotation frequency (circulation or "angular momentum") of the coordinate system. The angle $\varphi_0$ is fixed arbitrarily at the onset of symmetry breaking.
\subsection{The pure gravitational radiation case}
So far, we found that temporarily off-diagonal terms occurred in the perturbative approach of the Einstein scalar gauge field.
What remains unclear is if the breaking of the axially symmetry already appears in the vacuum case like in the vicinity of the black hole spacetime.
It is conjectured that the formation of primordial (Kerr-) black holes (and so quasars) happened in the early stages of the evolution of the universe before the stars were formed.
Therefore, consider the radiative Vaidya spacetime in Eddington-Finkelstein coordinates \footnote{This spacetime is also applied to describe the evaporation of a black hole by hawking radiation in a quantum mechanical way.}
\begin{equation}
ds^2=-\Bigl(1-\frac{2M(u)}{r}\Bigr)du^2-2dudr+r^2(d\theta^2+\sin^2\theta d\varphi^2),\label{3-12}
\end{equation}
which is the Schwarzschild  black hole spacetime  with $u=t-r-2M\log(\frac{r}{2M}-1)$.
Here we used $l_\mu l^\mu =0$. In the radiative coordinates, we have $x^1=u=\Theta({\bf x})$ and $ l_\mu =(1,0,0,0)$. From Eq.(\ref{3-6}) we obtain
\begin{equation}
h_{rr}=h_{r\theta}=h_{r\varphi}=0,\qquad h_{\varphi\varphi}=-\sin^2\theta h_{\theta\theta}\label{3-13}
\end{equation}
From the zero-order equations Eq.(\ref{3-7}) we obtain
\begin{eqnarray}
\ddot k_{rr}=0,\qquad \dot h_{\theta\theta}=r\partial_r\dot h_{\theta\theta},\qquad
\dot h_{\theta\varphi}=r\partial_r\dot h_{\theta\varphi}.\label{3-14}
\end{eqnarray}
So one writes
\begin{equation}
h_{\theta\theta}=r\alpha(u,\theta,\varphi,\xi),\quad h_{\theta\varphi}=r\beta(u,\theta,\varphi,\xi), \quad h_{\varphi\varphi}=-r\alpha\sin^2\theta.\label{3-15}
\end{equation}
Further, we have
\begin{eqnarray}
\ddot k_{r\theta}=\frac{1}{r}\Bigl(2\dot\alpha\cot\theta +\partial_\theta\dot\alpha+\frac{1}{\sin^2\theta}\partial_\varphi\dot\beta\Bigr),\label{3-16}
\end{eqnarray}
\begin{eqnarray}
\ddot k_{r\varphi}=\frac{1}{r}\Bigl(\dot\beta\cot\theta -\partial_\varphi\dot\alpha +\partial_\theta\dot\beta\Bigr),\label{3-17}
\end{eqnarray}
\begin{eqnarray}
\frac{dM}{du}=-\frac{\ddot k_{\phi\phi}+\sin^2\theta\ddot k_{\theta\theta}}{4\sin^2\theta}
-\frac{1}{2}r\dot h_{uu}-\frac{1}{4}\Bigl(\dot\alpha^2+\frac{\dot\beta^2}{\sin^2\theta}\Bigr)
+\frac{1}{4}\Bigl(\ddot{\alpha^2}+\frac{\ddot{\beta^2}}{\sin^2\theta}\Bigr).\label{3-18}
\end{eqnarray}
Not all the components of $h_{\mu\nu}$ and $k_{\mu\nu}$ are physical, so one needs some extra gauge conditions.
Suitable choice of $\alpha$ and $\beta$ (Choquet-Bruhat uses, for example, $ \alpha=0,  \beta=g(u)h(\xi)\sin\theta$), leads to a solution to second order which is in general not axially symmetric. We can integrate these zero order equations with respect to $\xi$. One obtains then some conditions on the background fields, because  terms like $\int \dot\alpha d\xi$ disappear.
From Eq.(\ref{3-18}), we obtain
\begin{equation}
\frac{dM}{du}=-\frac{1}{4\tau}\int_0^\tau\Bigl(\dot\alpha^2+\frac{\dot\beta^2}{\sin^2\theta}\Bigr)d\xi,\label{3-19}
\end{equation}
which is the back-reaction of the high-frequency disturbances on the mass $M$. $\tau$ is the period of $\dot h_{\mu\nu}$. This expression can be substituted back into Eq.(\ref{3-18}).
However, in the non-vacuum case, the right-hand side will also contain contributions from the matter fields.
In order to obtain propagation equations for $h_{\mu\nu}$ and $k_{\mu\nu}$, one  proceeds with the next order equation $R_{\mu\nu}^{(1)}=0$. First of all, Eq.(\ref{3-16}), (\ref{3-17}) are consistent with $R_{r\varphi}^{(1)}=0$ and $R_{r\theta}^{(1)}=0$.
Further, one obtains  propagation equations for $\alpha$ and $\beta$ and for some second order perturbations, such as $k_{\varphi\varphi}$. Moreover, the $(\varphi, \theta)$-dependent part of the PDE's for $\alpha$ and $\beta$
(say $A(\theta,\varphi), B(\theta,\varphi)$) can be separated (for the case $k_{\theta\varphi}\neq 0$): 
\begin{equation}
\partial_\varphi B+2\sin\theta\cos\theta A+\sin^2\theta\partial_\theta A=0,\label{3-20}
\end{equation}
\begin{eqnarray}
\sin^2\theta\partial_{\theta\theta}A+7\sin\theta\cos\theta\partial_\theta A+4\cot\theta\partial_\varphi B +2(5\cos^2\theta-1)A+2\partial_{\theta\varphi}B=0.\label{3-21}
\end{eqnarray}
A non-trivial simple solution is
\begin{equation}
A=\frac{\cos\theta (sin\varphi+\cos\varphi)}{sin^3\theta},\quad B=\frac{\sin\varphi-\cos\varphi}{\sin^2\theta}+G(\theta),\label{3-22}
\end{equation}
with $G(\theta)$ arbitrary. So the breaking of the spherically and axially symmetry is evident.
\section{Conclusions}\label{sec:concl}
There is clear new observational evidence for  the azimuthal alignment of the spin vectors of quasars in three new studied LQG. This research presents a new argument about the theoretical explanation of the axial symmetry breaking in a non-linear perturbation scheme considering a vacuum black hole spacetime in radiative coordinates. The recently discovered 13 billion years old quasar P172+18 powered by a supermassive black hole, is all the more reason to believe that the formation of these objects took place in the very early universe.

\section{Appendix: The Data}
The data of LQG U1.27 underlying this article are  gathered in Table 1 and 2.
\begin{table*}
\caption{Data for the LQG U1.27 (N=71) from NASA/IPAC  and SIMBAD. The successive columns represent: right ascension, declination, redshift, inclination, eccentricity, position angle, azimuthal angle and polar angle. }
\centering 
\begin{tabular}{c c c c c c c c c} 
\hline\hline 
U1.27 & RA & Dec & z & inc & ecc & PA(deg) & $\varphi$(rad) & $\theta$(rad) \\ [0.5ex]
\hline 
1 & 160.413150 & 14.591740 & 1.221 & .830 & .69 & 87 & $\pm$.298 / $\pm$.427 & .575 ($\pi$-.575) / 1.083 ($\pi$-1.083)\\ 
2 & 160.840103 & 14.600057 & 1.271 & .606 & .83 & 113 & .571 / .013 & .305 / .796 \\ 
3 & 161.128848 & 16.045854 & 1.233 & .526 & .87 & 45 & .034 / .782 & .619 / .102 \\ 
4 & 161.187666 & 15.317133 & 1.237 & .741 & .75 & 68 & .049 / .763 & .421 / .925 \\ 
5 & 161.335962 & 14.290068 & 1.270 & .586 & .84 & 63 & .063 / .678 & .751 / .275 \\ 
6 & 161.516893 & 14.044789 & 1.290 & .547 & .86 & 6 & .227 / .889 & -.155 / .263 \\ 
7 & 161.567993 & 16.753510 & 1.282 & .772 & .73 & 48 & .186 / 1.04 & .779 / .294 \\ 
8 & 161.601093 & 14.502540 & 1.372 & .660 & .80 & 60 & .007 / .775 & .322 / .792 \\ 
9 & 162.056815 & 16.480304 & 1.290 & .506 & .88 & 170 & .82 / .215 & .336 / -.167 \\ 
10 & 162.248981 & 12.889527 & 1.368 & .677 & .79 & 99 & .419 / .156 & .434 / .890 \\ 
11 & 162.344193 & 15.726700 & 1.263 & .772 & .73 & 42 & .257 / 1.05 & .258 / .699 \\ 
12 & 162.351272 & 15.698897 & 1.301 & .435 & .91 & 8 & .129 / .762 & -.190 / .306 \\ 
13 & 162.409269 & 21.808144 & 1.235 & .526 & .87 & 33 & .127 / .849 & -.067 / .612 \\ 
14 & 162.423677 & 15.306867 & 1.341 & .676 & .79 & 3 & .379 / 1.00 & -.175 / .240 \\ 
15 & 162.449082 & 16.371282 & 1.300 & .566 & .85 & 142 & .745 / .225 & .080 / .589 \\ 
16 & 162.505093 & 15.565017 & 1.255 & .357 & .94 & 17 & .038 / .671 & -.154 / .357 \\ 
17 & 162.676146 & 16.015594 & 1.269 & .771 & .73 & 125 & .744 / .340 & .843 / .359 \\
18 & 162.767371 & 16.316926 & 1.253 & .287 & .96 & 170 & .592 / .004 & .322 / -.224 \\ 
19 & 162.820881 & 13.193341 & 1.337 & .483 & .89 & 102 & .400 / .175 & .700 / .243 \\ 
20 & 162.831696 & 14.436524 & 1.315 & .659 & .80 & 54 & .084 / .812 & .743 / .287 \\ 
21 & 162.845783 & 11.981222 & 1.309 & .437 & .91 & 159 & .707 / .132 & .344 / -.039 \\ 
22 & 162.857213 & 12.796204 & 1.283 & .641 & .81 & 139 & .779 / .275 & .595 / .207 \\ 
23 & 162.884258 & 14.937553 & 1.367 & .756 & .74 & 94 & .354 / .211 & 1.014 / .494 \\ 
24 & 162.918357 & 20.655882 & 1.174 & .567 & .85 & 49 & .062 / .799 & .743 / .082 \\ 
25 & 162.937035 & 12.974699 & 1.316 & .461 & .90 & 110 & .455 / .107 & .654 / .207 \\ 
26 & 163.041780 & 16.928818 & 1.339 & .436 & .91 & 59 & .077 / .576 & .656 / .083 \\ 
27 & 163.092251 & 12.515036 & 1.316 & .355 & .94 & 21 & .035 / .647 & -.082 / .331 \\ 
28 & 163.098712 & 14.090468 & 1.256 & .622 & .82 & 101 & .415 / .127 & .365 / .852 \\ 
29 & 163.100377 & 20.776172 & 1.203 & .504 & .88 & 25 & .160 / .827 & -.120 / .525 \\ 
30 & 163.190868& 13.682636 & 1.356 & .482 & .89 & 60 &  .057 / .590 & .181 / .643 \\ 
31 & 163.238226 & 10.992647 & 1.266 & .483 & .89 & 89 & .285 / .305 & .291 / .674 \\ 
32 & 163.242363 & 20.284854 & 1.253 & .356 & .94 & 41 & .026 / .611 & -.111 / .569 \\ 
33 & 163.552829 & 14.959790 & 1.231 & .678 & .79 & 120 & .625 / .184 & .330 / .812 \\
34 & 163.591268 & 21.358676 & 1.257 & .606 & .83 & 120 & .579 / .164 & .161 / .862 \\ 
35 & 163.648532 & 10.304548 & 1.260 & .385 & .93 & 16 & .084 / .671 & -.064 / .271 \\ 
36 & 163.677979 & 10.722398 & 1.335 & .725 & .76 & 144 & .872 / .393 & .247 / .550 \\ 
37 & 163.694730 & 19.952953 & 1.220 & .566 & .85 & 20 & .246 / .889 & -.116 / .478 \\ 
38 & 163.845984 & 13.102997 & 1.358 & .588 & .84 & 11 & .295 / .889 & .296 / -.085 \\ 
39 & 163.854942 & 19.298998 & 1.201 & .773 & .73 & 146 & .906 / .531 & .651 / .133 \\ 
40 & 163.857047 & 11.617507 & 1.293 & .527 & .87 & 119 & .538 / .027 & .650 / .260 \\ 
[1ex] 
\hline 
\end{tabular}
\label{table:nonlin} 
\end{table*}
\begin{table*}
\caption{continue }
\centering 
\begin{tabular}{c c c c c c c c c} 
\hline\hline 
cont. & RA & Dec & z & inc & ecc & PA(deg) & $\varphi$(rad) & $\theta$(rad) \\ [0.5ex]
\hline 
41 & 163.924334 & 11.298387 & 1.331 & .845 & .68 & 88 & .249 / .335 & .648 / 1.042 \\ 
42 & 163.984288 & 18.788475 & 1.277 & .355 & .94 & 87 & .262 / .305 & .027 / .683 \\ 
43 & 164.046989 & 17.140997 & 1.344 & .845 & .68 & 72 & .012 / .780 & .505 / 1.066 \\ 
44 & 164.091276 & 14.566966 & 1.243 & .207 & .98 & 129 & .409 / .138 & .412 / -.091 \\ 
45 & 164.156218 & 15.013202 & 1.371 & .787 & .72 & 109 & .541 / .147 & .979 / .483 \\ 
46 & 164.158278 & 10.052025 & 1.273 & .435 & .91 & 128 & .544 / .023 & .507 / .170 \\ 
47 & 164.230688 & 14.829509 & 1.229 & .625 & .82 & 133 & .696 / .257 & .208 / .671 \\ 
48 & 164.308435 & 18.798158 & 1.285 & .382 & .91 & 152 & .614 / .106 & -.133 / .484 \\ 
49 & 164.521231 & 20.061417 & 1.273 & .587 & .94 & 72 & .095 / .549 & .211 / .895 \\
50 & 164.633396 & 17.082247 & 1.286 & .462 & .90 & 17 & .175 / .750 & -.139 / .398 \\ 
51 & 164.668752 & 17.904320 & 1.269 & .527 & .87 & 50 & .062 / .700 & .101 / .684 \\ 
52 & 164.730550 &  8.230752 & 1.246 & .677 & .79 & 8 & .401 / .953 & -.025 / .199 \\ 
53 & 164.869078 & 16.782794 & 1.300 & .623 & .82 & 116 & .533 / .122 & .272 / .829 \\ 
54 & 165.025092 &  9.444098 & 1.252 & .567 & .85 & 65 & .019 / .556 & .348 / .666 \\ 
55 & 165.070384 & 19.606880 & 1.240 & .832 & .69 & 114 & .597 / .372 & .423 / 1.040 \\ 
56 & 165.166670 & 16.952878 & 1.300 & .802 & .71 & 127 & .739 / .454 & .354 / .851 \\ 
57 & 165.452792 &  8.368669 & 1.196 & .413 & .92 & 9 & .153 / .671 & -.071 / .197 \\ 
58 & 165.676652 &  8.655867 & 1.240 & .411 & .92 & 60 & .046 / .481 & .205 / .500 \\ 
59 & 166.268588 &  8.759810 & 1.241 & .566 & .85 & 57 & .072 / .606 & .322 / .610 \\ 
60 & 166.589189 &  8.686458 & 1.244 & .548 & .86 & 53 & .098 / .618 & .286 / .571 \\ 
61 & 166.902560 &  9.020778 & 1.228 & 1.02 & .55 & 87 & .161 / .347 & .860 / 1.175 \\ 
62 & 166.935900 &  9.924171 & 1.225 & .383 & .93 & 14 & .143 / .613 & -.071 / .251 \\ 
63 & 167.532914 & 10.802888 & 1.211 & .248 & .97 & 129 & .374 / .052 & .006 / .378 \\ 
64 & 167.539961 &  7.868564 & 1.208 & .740 & .75 & 10 & .508 / .967 & .015 / .219 \\ 
65 & 168.567405 & 10.390996 & 1.210 & .693 & .78 & 151 & .803 / .472 & .168 / .460 \\
66 & 168.938775 &  8.249943 & 1.194 & .547 & .86 & 106 & .349 / .012 & .381 / .665 \\ 
67 & 169.508801 & 10.550690 & 1.215 & .526 & .87 & 134 & .548 / .234 & .196 / .540 \\ 
68 & 169.596744 &  9.084715 & 1.197 & .787 & .72 & 98 & .303 / .018 & .620 / .934 \\ 
69 & 170.081761 &  8.984773 & 1.229 & .641 & .81 & 56 & .193 / .613 & .373 / .662 \\ 
70 & 170.246993 & 10.185907 & 1.208 & .384 & .93 & 77 & .085 / .271 & .196 / .551 \\ 
71 & 170.290711 &  7.999635 & 1.141 & .461 & .90 & 93 & .195 / .143 & .320 / .599 \\
[1ex] 
\hline 
\end{tabular}
\label{table:nonlin} 
\end{table*} 
\bibliography{slagter}{}
\bibliographystyle{aasjournal}
\end{document}